\documentclass[a4paper,11pt]{article}
\usepackage{pos}

\usepackage{tikz,graphicx,booktabs,multirow,array,microtype,mathtools,float}

\usepackage{comment}

\usepackage{centernot}

\graphicspath{{figures/}}
\usepackage{enumerate}   
\extrafloats{200}
\usepackage{anyfontsize}
\usepackage[shortlabels]{enumitem}

\usepackage{amsfonts}
\AtBeginDocument{
\heavyrulewidth=.08em
\lightrulewidth=.05em
\cmidrulewidth=.03em
\belowrulesep=.65ex
\belowbottomsep=0pt
\aboverulesep=.4ex
\abovetopsep=0pt
\cmidrulesep=\doublerulesep
\cmidrulekern=.5em
\defaultaddspace=.5em
}

\newcommand{\langl}{\begin{picture}(4.5,7)
\put(1.1,2.5){\rotatebox{60}{\line(1,0){5.5}}}
\put(1.1,2.5){\rotatebox{300}{\line(1,0){5.5}}}
\end{picture}}
\newcommand{\rangl}{\begin{picture}(4.5,7)
\put(.9,2.5){\rotatebox{120}{\line(1,0){5.5}}}
\put(.9,2.5){\rotatebox{240}{\line(1,0){5.5}}}
\end{picture}}

\usepackage{amsmath} 
\usepackage{dsfont}  
\usepackage{bm}      

\def\beq{\begin{equation}}
\def\eeq{\end{equation}}
\def\beqs#1\eeqs{\beq\begin{split} #1 \end{split}\eeq}

\long\def\comment#1{}

\title{Electric Polarizability of Charged Kaons from Lattice QCD Four-Point Functions}

\author*[a]{Shayan Nadeem}
\author[a]{Walter Wilcox}
\author[b]{Frank X. Lee}

\affiliation[a]{Department of Physics and Astronomy, Baylor University,\\ Waco, Texas 76798, USA}

\affiliation[b]{Physics Department, The George Washington University,\\
 Washington, District of Columbia 20052, USA}

\emailAdd{shayan\_nadeem1@baylor.edu}
\emailAdd{walter\_wilcox@baylor.edu}
\emailAdd{fxlee@gwu.edu}

\abstract{We study the electric polarizability of a charged kaon from four-point functions in lattice QCD as
an alternative to the background field method. Lattice four-point correlation functions are constructed from quark and gluon fields to be used in Monte Carlo simulations. The elastic form factor
(charge radius) is needed in the method which can be obtained from the same four-point functions
at large current separations. Preliminary results from the connected quark-line diagrams are
presented.}

\FullConference{The 41st International Symposium on Lattice Field Theory (LATTICE2024)\\
 28 July - 3 August 2024\\
Liverpool, UK\\}

\begin{document}
\maketitle

\section{\label{sec:intro}Introduction}

{The study of electromagnetic polarizabilities is a long-standing pursuit in hadronic physics, and its investigation within lattice QCD presents both intriguing opportunities and considerable challenges. Traditionally, the background field method has been the go-to technique for computing polarizabilities, providing reliable results for neutral hadrons. This approach has seen widespread application in various lattice studies~\cite{Fiebig:1988en,Lujan:2016ffj, Freeman:2014kka}. However, the situation becomes considerably more complicated when dealing with charged particles. In this case, the problem is twofold: the quenching of the external electromagnetic field and the fact that charged hadrons, when placed in an external field, will experience effects like the formation of Landau levels in a magnetic field. These challenges have limited the number of lattice QCD calculations for charged hadrons, with much of the focus remaining on neutral mesons, such as the pion.

In this work, we employ the four-point functions method—an approach that, while not new, has not received as much attention in the context of polarizabilities. Four-point correlation functions have been used to study a variety of hadronic properties~\cite{Liang:2019frk}, but their potential for extracting polarizabilities has only recently been appreciated. To our knowledge, there have been two notable attempts in the past, one using position-space methods~\cite{BURKARDT1995441} and the other in momentum space~\cite{Wilcox:1996vx}. 

The four-point function method is ideally suited for studying charged hadrons. By directly incorporating the effects of the hadronic structure in a manner that avoids many of the pitfalls of the background field approach, this method holds promise for more precise and robust calculations. Our goal is to present a detailed study of the electric polarizability of the charged kaon, using lattice QCD four-point functions, and to compare our results with those obtained from other approaches.}

\section{Charged Kaon}
\label{sec:kaon}

\begin{figure}[h]
\begin{center}
\includegraphics[scale=1.0]{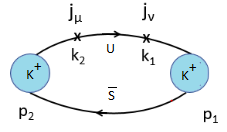}
\caption{Pictorial representation of the four-point function in Eq.\eqref{eq:kaon4pt} for $K^+$. 
Time flows from right to left and the four-momentum conservation is $p_2 +k_2 = k_1+p_1$.}
\label{fig:4pdiag}
\end{center}
\end{figure}

\begin{figure}[h]
\begin{center}
\includegraphics[scale=0.35]{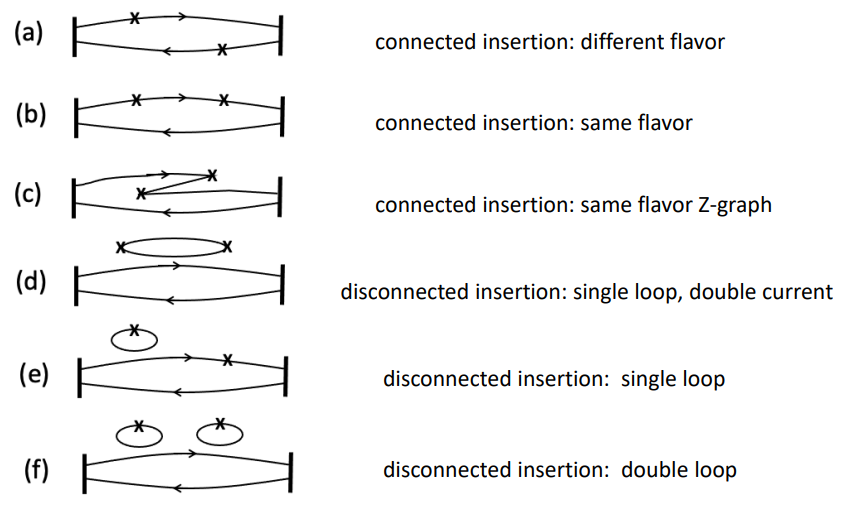}
\caption{Skeleton diagrams of a four-point function contributing
to polarizabilities of a meson: (a) connected insertion: different
flavor, (b) connected insertion: same flavor, (c) connected insertion: same flavor Z-graph, (d) disconnected insertion: single
loop, double current, (e) disconnected insertion: single loop,
(f) disconnected insertion: double loop. In each diagram, flavor
permutations are assumed as well as gluon lines that connect the
quark lines. The zero-momentum pion interpolating fields are
represented by vertical bars (wall sources). Time flows from right
to left.}
\label{fig:diagram-4pt1}
\end{center}
\end{figure}

In Ref.~\cite{pion4p} a formula for the electric polarizability of the charged pion is derived. For the kaon, the formula is the same except for the replacement of the pion mass with the kaon mass,
\begin{equation}
\alpha^K_E= {\alpha \langl  r_E^2\rangl \over 3m_{K}}+\lim_{\bm q\to 0}{2\alpha  \over \bm q^{\,2}} \int_{0}^\infty d t \bigg[Q_{44}(\bm q,t) -Q^{elas}_{44}(\bm q,t) \bigg].
\label{eq:kaon4pt}
\end{equation}
Here, $\alpha=1/137$ represents the fine structure constant. The first term in the equation includes the charge radius and the kaon mass, which we will refer to as the elastic contribution. The second term results from subtracting the elastic contribution $Q_{44}^{elas}$ from the total, and we will call this the inelastic contribution. This formula is applied in discrete Euclidean spacetime, though we retain a continuous Euclidean time axis for ease of notation. Special kinematics, known as the zero-momentum Breit frame, are used in the formula to simulate low-energy Compton scattering. The process is illustrated in Fig.~\ref{fig:4pdiag}. Part of evaluating the four-point function is to evaluate the topologically distinct quark-line diagrams. These diagrams are shown in Fig.~\ref{fig:diagram-4pt1}. The raw correlation functions can be found in Ref.~\cite{pion4p}.
\section{Simulation details and results}
\label{results}
{It is worth mentioning that our current results have some limitations. Firstly, we use 99 configurations for our analysis. This explains the comparatively large error bars. We are currently working towards performing an analysis using 500 configurations. Secondly, as a proof-of-principle simulation, we use quenched Wilson fermions.}
\subsection{Raw correlation functions}
{We present in Fig.~\ref{fig:diag_abc_seperate} the raw normalized four-point functions $Q_{44}$ at two
different values of momentum $\bm q$ and at $m_{\pi}=600 MeV$. These plots exhibit several interesting features. First, the point where $t_1=t_2$ behaves regularly in diagram (a) but yields irregular results in diagrams (b) and (c) for all values of $\bm q$. This irregularity corresponds to the contact term discussed in Ref.~\cite{pion4p}, and we exclude this point from our analysis. Second, the results around $t_1=18$ in diagrams (b) and (c) are mirror images of each other, which is due to the two different time orderings of the same diagram. In principle, this symmetry could be leveraged to reduce the computational cost of simulations. However, in this study, we computed all three diagrams separately.
\begin{figure}[H]
\begin{center}
\includegraphics[scale=0.45]{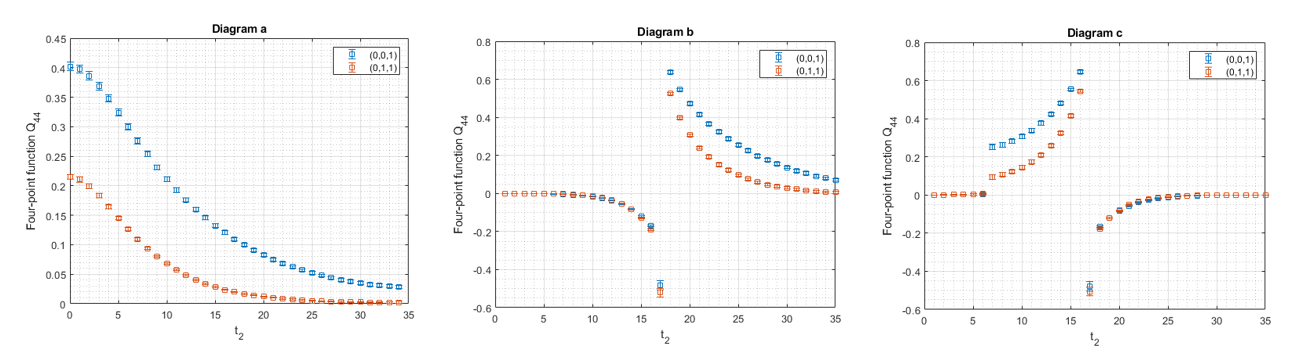}
\caption{Normalized four-point functions from the connected diagrams as a function of current separation at $m_{\pi}=600 MeV$.}
\label{fig:diag abc seperate}
\end{center}
\end{figure}
%
%

\subsection{Elastic form factor}
{
The formula for electric polarizability in Eq.\eqref{eq:kaon4pt} includes the charge radius $r_E$ and the elastic contribution $Q^{elas}_{44}$, both of which can be determined from the long-time behavior of the four-point functions $Q_{44}$.  According to the following equation given in Ref.~\cite{pion4p},
\beq
Q^{elas}_{44}(\bm q,t)= {(E_K+m_K)^2\over 4 E_K m_K} F^2_K(\bm q^2) \, e^{-a(E_K-m_K) t}.
\label{eq:Q44elas}
\eeq
$Q^{elas}_{44}$ is expected to follow a single-exponential behavior with a decay rate of $E_K-m_K$. The form factor $F_K$ is embedded in the amplitude of this decay. As discussed in Ref.~\cite{pion4p}, diagrams $a$ and $b$ display the expected decay, while diagram $c$ does not. For the elastic contribution, we can omit
\begin{figure}[H]
\begin{center}
\includegraphics[scale=0.55]{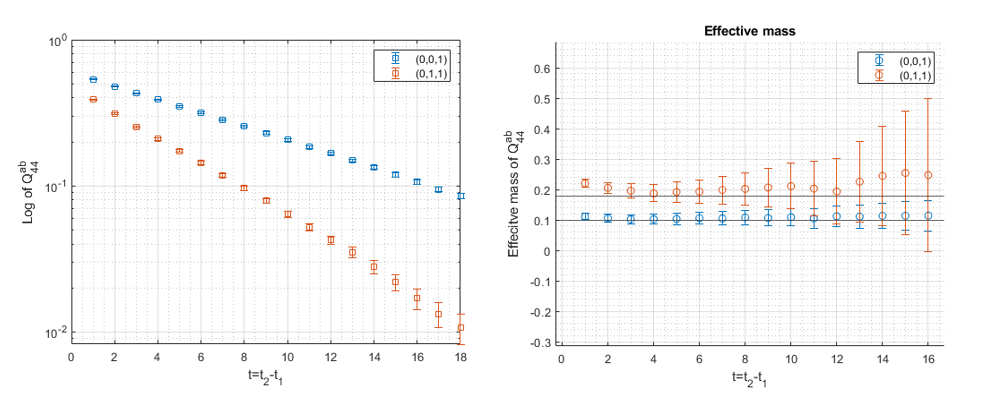}
\caption{Normalized four-point functions from diagrams $a$ and $b$ in log plot and their effective mass functions at different values of $\bm q$ and $m_\pi=600$ MeV. They are plotted as functions of time separation $t=t_2-t_1$ between the two currents relative to fixed $t_1=18$. The horizontal gridlines in the effective mass are $E_K-m_K$ using continuum dispersion relation for $E_K$ with measured $m_K$. These functions are used to extract the elastic contributions $Q^{elas}_{44}$.}
\label{fig:EMms2}
\end{center}
\end{figure}
\noindent diagram $c$ and concentrate on diagrams $a$ and $b$, which enhances the form factor analysis by removing the inelastic 'contamination' from diagram $c$, serving as an optimization in the analysis.

Fig.~\ref{fig:EMms2} provides an example of the four-point functions $Q^{ab}_{44}$, including only diagrams $a$ and $b$, along with their effective mass functions. We focus on the signal region between $t_1$ and $t_3$, plotting them as a function of the time separation $t=t_2-t_1$ between the two currents. The $t=0$ point is excluded from the analysis due to contact terms, as mentioned earlier. There is a region where the effective mass functions align with the $E_K-m_K$ gridlines, indicating that $Q^{ab}_{44}$ is primarily governed by elastic contributions. This agreement is more pronounced at lower momentum values. At larger times, the signal becomes noisy, particularly at higher momentum.

To address potential deviations from the continuum dispersion relation, we fit the functional form of $Q^{elas}_{44}$ in Eq.\eqref{eq:Q44elas}, treating ${F_K, E_K}$ as free parameters while keeping $m_K$ fixed at its measured value from two-point functions. After the form factor data are obtained, we fit them to the monopole form,
\beq
F_\pi(\bm q^2) = \frac{1}{1+\bm q^2/m_V^2},
\label{eq:vmd}
\eeq
which is the well-known vector meson dominance (VMD)  commonly considered in pion form factor studies. We use the monopole form because of the availability of just two momenta. As we move on with our analysis, we will be working with data for four momenta, at which point we will be using the z-expansion parametrization ~\cite{zexpansion} for a better fit. The results are illustrated in Fig.~\ref{fig:ff}.
\begin{figure}[h!]
\centering
\includegraphics[scale=0.55]{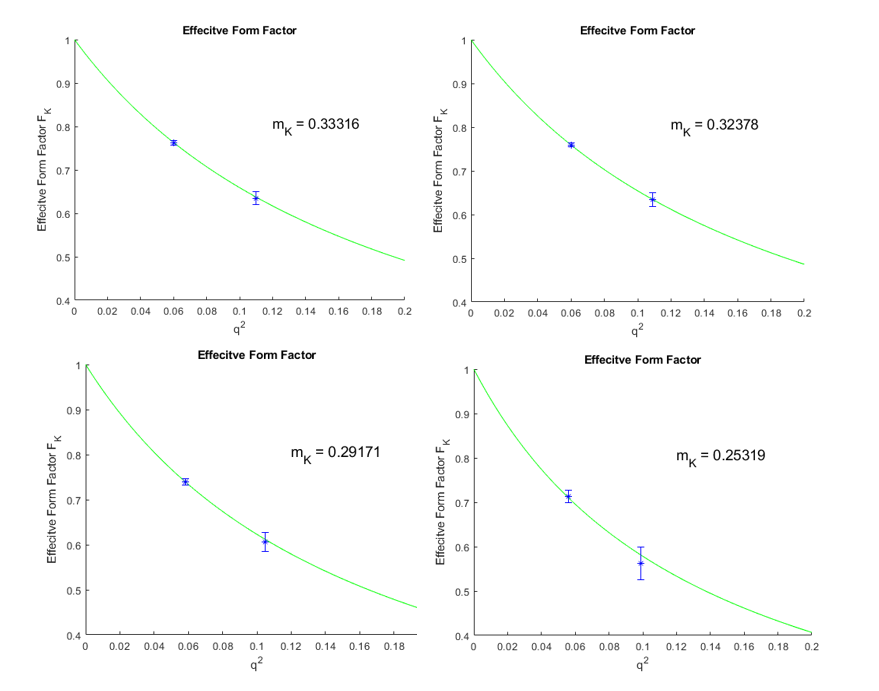}
\caption{
Pion elastic form factors extracted from four-point functions. The blue data points are the measured values. The green solid line is a fit to the monopole form in Eq.~\eqref{eq:vmd}. $m_K$ and $q^2$ values in lattice units.
}
\label{fig:ff}
\end{figure}
\noindent Once the functional form of form factor is determined, the charge radius is  obtained by 
\beq
\langl  r_E^2\rangl= -6 \frac{dF_K (\bm q^2)}{d \bm q^2 } \Big|_{\bm q^2\to 0}.
\label{eq:re2}
\eeq
}
\subsection{Electric polarizability}
{After determining the elastic contribution $Q_{44}^{elas}$, we now focus on the inelastic part of $\alpha_E$ from Eq.\eqref{eq:kaon4pt}. In Fig.~\ref{fig:QQ}, we present the total contribution $Q_{44}$ (from all three diagrams) and $Q_{44}^{elas}$ as functions of the current separation $t=t_2-t_1$, using $m_\pi=600$ MeV as an example. The graphs for other pion masses are similar. Although $Q_{44}^{elas}$ is derived from the large-time region, the subtraction is applied across the entire region based on the functional form in Eq.\eqref{eq:Q44elas}. Most contributions arise from the small-time region where inelastic effects are more prominent. We observe that $Q_{44}^{elas}$ consistently exceeds $Q_{44}$, implying a negative inelastic term in the formula. The time integral corresponds to the negative of the area between the two curves.
\begin{figure}[h!]
\begin{center}
\includegraphics[scale=0.8]{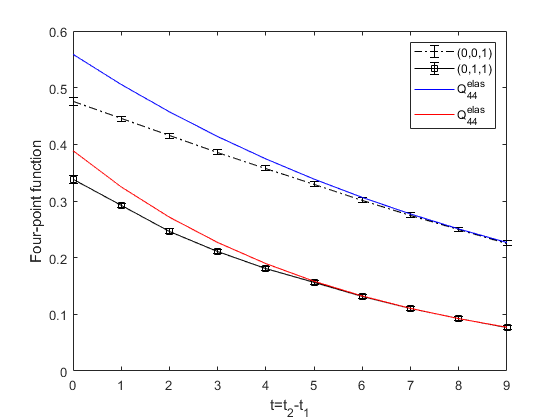}
\caption{
Total $Q_{44}$ and elastic $Q_{44}^{elas}$  at different values of $\bm q$ at $m_\pi=600$ MeV.
 The area between the curves, $(1/a)\int dt \big[ Q_{44}(\bm q,t)-Q_{44}^{elas}(\bm q,t)\big]$,  is the dimensionless signal contributing to polarizability.
}
\label{fig:QQ}
\end{center}
\end{figure}
Notably, the curves include the $t=0$ point, which contains unphysical contributions in $Q_{44}$, as previously mentioned. Typically, we would exclude this point and start the integral from $t=1$. However, the area between $t=0$ and $t=1$ constitutes the largest portion of the integral. To account for this, we linearly extrapolated the $Q_{44}$ term back to $t=0$ using the values at $t=1$ and $t=2$. This introduces a systematic effect of order $O(a^2)$, as the error itself is of order $O(a)$. This effect diminishes as the continuum limit is approached, with the area shrinking to zero. Including this point in $Q_{44}^{elas}$ using its functional form poses no issue.

The inelastic term is constructed by multiplying ${2\alpha / \bm q^{,2}}$ by the time integral; this entire term is a function of momentum. Since $\alpha_E$ is a static property, we smoothly extrapolate it to $\bm q^2=0$. Due to the limitation of having only two momentum values, we use a linear fit for this extrapolation. Finally, we combine the two terms in the formula from Eq.\eqref{eq:kaon4pt} to calculate $\alpha_E$ in physical units.

To examine the trend at smaller pion masses, we will take the total values for $\alpha_E$ at four pion masses and perform a smooth extrapolation to the physical point. Currently, we only display a connected line between data points. 
%
%
%
\begin{figure}[t!]
\begin{center}
\includegraphics[scale=0.79]{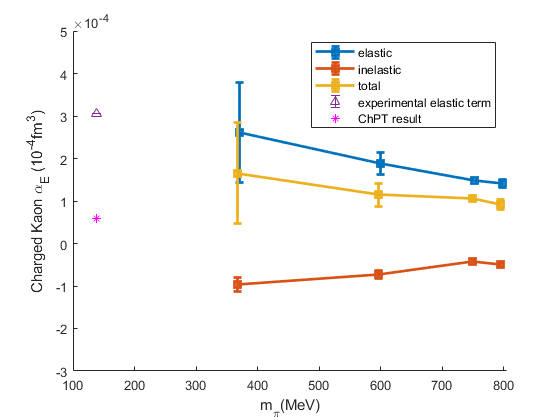}
\caption{
Kaon mass dependence of electric polarizability  of a charged kaon from  four-point functions in lattice QCD.
Elastic and inelastic contributions correspond to the two terms in the formula in Eq.\eqref{eq:kaon4pt}. 
Magenta triangle is the experimental PDG value for the elastic term.
Pink star is the ChPT value for the kaon electric polarizability. The elastic data points are manually shifted to the right for better visibility.}
\label{fig:finalpolar}
\end{center}
\end{figure}
The results are summarized in Fig.~\ref{fig:finalpolar}. At the pion masses studied, our lattice results reveal a distinct pattern for electric polarizability: the elastic term contributes positively, while the inelastic term contributes negatively but with a smaller magnitude. This partial cancellation results in a positive total value. The cancellation appears to persist as we approach the physical point, though it is less quantitatively conclusive, as indicated by the uncertainty bands from the extrapolations. This underscores the importance of investigating smaller pion masses in future simulations.
The results in Fig.~\ref{fig:finalpolar} show a similar trend to the results in Ref.~\cite{pion4p} for the pion. It can be seen that the electric polarizability increases with decreasing pion mass. It also seems that an extrapolation to the physical point will result in a value higher than the ChPT result, which is what we see for the pion as well. 
}
\subsection{Conclusions}
{In this study, we have explored an alternative approach to calculating the electric polarizability of the charged kaon using four-point functions in lattice QCD.  By leveraging four-point correlation functions, we overcome challenges that arise from electro-quenching and the complexities of charged hadrons in external electromagnetic fields, as discussed in previous works~\cite{Wilcox:1996vx,BURKARDT1995441}. 

Our results highlight the separation of the elastic and inelastic contributions to the kaon's electric polarizability. The elastic term contributes positively, while the inelastic term introduces a smaller negative contribution, leading to a cancellation between the two terms. These findings are consistent with earlier studies of charged hadrons~\cite{Engelhardt:2007ub, Lee:2005dq}.

The relatively small number of configurations (99) used in this study introduces statistical uncertainties. We are working towards using a larger configuration set (500) to help reduce these uncertainties significantly. Moreover, the current simulation utilizes quenched Wilson fermions, which have known limitations, as evidenced in earlier studies~\cite{Alexandru:2008sj, Freeman:2014kka}. Future work will involve using dynamical fermions, which will better reflect the physical quark content and improve the reliability of the results.

As we move forward, we will also be using data for 4 momenta instead of 2, and apply more sophisticated fitting methods, such as the z-expansion~\cite{zexpansion}, to enhance the accuracy of our results. Our analysis also suggests that a more detailed exploration of smaller pion masses will be important for refining the extrapolation to the physical point, a task that has been undertaken in other works studying meson polarizabilities~\cite{Bignell:2020xkf, Lujan:2016ffj}.}

\acknowledgments
We would like to acknowledge support from the Baylor College of Arts and Sciences Summer Research Award (SRA) program. This work was supported in part by DOE Grant No.~DE-FG02-95ER40907. The authors acknowledge the Texas Advanced Computing Center (TACC) at The University of Texas at Austin for providing computational resources that have contributed to the research results reported within this paper.

\bibliographystyle{JHEP}
\bibliography{x4ptfun}

\providecommand{\href}[2]{#2}\begingroup\raggedright\begin{thebibliography}{10}

\bibitem{Fiebig:1988en}
H.R.~Fiebig, W.~Wilcox and R.M.~Woloshyn, \emph{{A Study of Hadron Electric Polarizability in Quenched Lattice QCD}}, \href{https://doi.org/10.1016/0550-3213(89)90180-6}{\emph{Nucl. Phys. B} {\bfseries 324} (1989) 47}.

\bibitem{Lujan:2016ffj}
M.~Lujan, A.~Alexandru, W.~Freeman and F.X.~Lee, \emph{{Finite volume effects on the electric polarizability of neutral hadrons in lattice QCD}}, \href{https://doi.org/10.1103/PhysRevD.94.074506}{\emph{Phys. Rev.} {\bfseries D94} (2016) 074506} [\href{https://arxiv.org/abs/1606.07928}{{\ttfamily 1606.07928}}].

\bibitem{Freeman:2014kka}
W.~Freeman, A.~Alexandru, M.~Lujan and F.X.~Lee, \emph{{Sea quark contributions to the electric polarizability of hadrons}}, \href{https://doi.org/10.1103/PhysRevD.90.054507}{\emph{Phys. Rev. D} {\bfseries 90} (2014) 054507} [\href{https://arxiv.org/abs/1407.2687}{{\ttfamily 1407.2687}}].

\bibitem{Liang:2019frk}
{\scshape XQCD} collaboration, \emph{{Towards the nucleon hadronic tensor from lattice QCD}}, \href{https://doi.org/10.1103/PhysRevD.101.114503}{\emph{Phys. Rev. D} {\bfseries 101} (2020) 114503} [\href{https://arxiv.org/abs/1906.05312}{{\ttfamily 1906.05312}}].

\bibitem{BURKARDT1995441}
M.~Burkardt, J.~Grandy and J.~Negele, \emph{{Calculation and Interpretation of Hadron Correlation Functions in Lattice QCD}}, \href{https://doi.org/https://doi.org/10.1006/aphy.1995.1026}{\emph{Annals Phys.} {\bfseries 238} (1995) 441} [\href{https://arxiv.org/abs/hep-lat/9406009}{{\ttfamily hep-lat/9406009}}].

\bibitem{Wilcox:1996vx}
W.~Wilcox, \emph{{Lattice charge overlap. 2: Aspects of charged pion polarizability}}, \href{https://doi.org/10.1006/aphy.1996.5649}{\emph{Annals Phys.} {\bfseries 255} (1997) 60} [\href{https://arxiv.org/abs/hep-lat/9606019}{{\ttfamily hep-lat/9606019}}].

\bibitem{pion4p}
F.X.~Lee, A.~Alexandru, C.~Culver and W.~Wilcox, \emph{{Charged pion electric polarizability from four-point functions in lattice QCD}}, \href{https://doi.org/10.1103/PhysRevD.108.014512}{\emph{Phys. Rev. D} {\bfseries 108} (2023) 014512} [\href{https://arxiv.org/abs/2301.05200}{{\ttfamily 2301.05200}}].

\bibitem{zexpansion}
G.~Lee, J.R.~Arrington and R.J.~Hill, \emph{Extraction of the proton radius from electron-proton scattering data}, \href{https://doi.org/10.1103/PhysRevD.92.013013}{\emph{Phys. Rev. D} {\bfseries 92} (2015) 013013} [\href{https://arxiv.org/abs/1505.01489}{{\ttfamily 1505.01489}}].

\bibitem{Engelhardt:2007ub}
M.~Engelhardt, \emph{{Neutron electric polarizability from unquenched lattice QCD using the background field approach}}, \href{https://doi.org/10.1103/PhysRevD.76.114502}{\emph{Phys. Rev. D} {\bfseries 76} (2007) 114502} [\href{https://arxiv.org/abs/0706.3919}{{\ttfamily 0706.3919}}].

\bibitem{Lee:2005dq}
F.X.~Lee, L.~Zhou, W.~Wilcox and J.C.~Christensen, \emph{{Magnetic polarizability of hadrons from lattice QCD in the background field method}}, \href{https://doi.org/10.1103/PhysRevD.73.034503}{\emph{Phys. Rev. D} {\bfseries 73} (2006) 034503} [\href{https://arxiv.org/abs/hep-lat/0509065}{{\ttfamily hep-lat/0509065}}].

\bibitem{Alexandru:2008sj}
A.~Alexandru and F.X.~Lee, \emph{{The Background field method on the lattice}}, \href{https://doi.org/10.22323/1.066.0145}{\emph{PoS} {\bfseries LATTICE2008} (2008) 145} [\href{https://arxiv.org/abs/0810.2833}{{\ttfamily 0810.2833}}].

\bibitem{Bignell:2020xkf}
R.~Bignell, W.~Kamleh and D.~Leinweber, \emph{{Magnetic polarizability of the nucleon using a Laplacian mode projection}}, \href{https://doi.org/10.1103/PhysRevD.101.094502}{\emph{Phys. Rev. D} {\bfseries 101} (2020) 094502} [\href{https://arxiv.org/abs/2002.07915}{{\ttfamily 2002.07915}}].

\end{thebibliography}\endgroup
%

\end{document}